\documentclass[prl,aps,showpacs,twocolumn]{revtex4}%
\usepackage{graphicx}
\usepackage{amsmath}
\usepackage{amsfonts}
\usepackage{amssymb}%
\providecommand{\U}[1]{\protect\rule{.1in}{.1in}}
\begin{document}
\title{Botanical Ratchets}
\author{I.M. Kuli\'{c}$^{1}$, M. Mani$^{1}$, H. Mohrbach$^{2}$, R. Thaokar$^{3}$ and
L. Mahadevan$^{1,4}$ }
\affiliation{$^{1}$Engineering and Applied Sciences, Harvard University, Cambridge, MA
02138, USA }
\affiliation{$^{2}$ Physique Moleculaire et des Collisions, Universite Paul Verlaine-57012
Metz, France, }
\affiliation{$^{3}$ Chemical Engineering, IIT Bombay, Mumbai 400076 India, }
\affiliation{$^{4}$ Organismic and Evolutionary Biology, Harvard University, Cambridge, MA
02138, USA}


\begin{abstract}
Ratcheting surfaces are a common motif in nature and appear in plant awns and
grasses. They are known to profer selective advantages for seed dispersion and
burial. In two simple model experiments we show that these anisotropically
toothed surfaces naturally serve as motion rectifiers and generically move in
a unidirectional manner when subjected to temporally and spatially symmetric
excitations of various origins. Using a combination of theory and experiment
we show that a linear relation between awn length and ratchet efficiency holds
under biologically relevant conditions. Thus, grass awns efficiently transform
non-equilibrium environmental stresses into useful work and directed motion
using their length as a fluctuation amplifier, yielding a selective advantage
to these organelles in many plant species.

\end{abstract}
\maketitle


Motion rectification by ratchets arises repeatedly as a theme with variations
in nature to technology on a range of scales, from the molecular
\cite{Feynman}-\cite{Kulic} to the macroscopic \cite{BugBerne}%
-\cite{Mahadevan} in both deterministic and stochastic systems. A key
ingredient in all ratchets is the presence of a static or dynamic asymmetry
that serves to rectify motion, with many convergent solutions seen in nature.
For example, ratchet and pawl mechanisms are basic mechanical components in
many man-made gadgets and machines. They serve to rectify linear or rotational
motions by using a mechanical anisotropy that often takes the form of an
asymmetric saw-tooth that allows the ratchet to slip in one direction and not
in the opposite direction. Their microscopic (thermal) analogs have been
intensely studied \cite{Feynman}-\cite{Kulic} over the past decades in context
of molecular motors \cite{Ratchet}, where the physical sawtooth is replaced by
a potential with a similar property that is periodically switched on and off.
On the macroscopic scale as well, nature has stumbled on ratchets repeatedly,
and a host of examples exist in the plant world in particular
\cite{OldPaperOnStipa}-\cite{Elbaum}, where they might profer selective
advantages for seed dispersion and burial. Indeed, the fact that different
plant families show the same robust and efficient solution for the dispersion
and self burial problem suggests that this is yet another example where a
simple physical principle often leads to convergent evolution. Here we use the
particular example of the wild foxtail \textit{Hordeum murinum} to study
macroscopic natural ratchets that rectify deformation induced by swelling,
shrinking, and other environmental factors into unidirectional motion that
serves important functional purposes, from the burial of seeds to their
transport by animals.

\begin{figure}[ptb]
\includegraphics*[width=8.5cm]{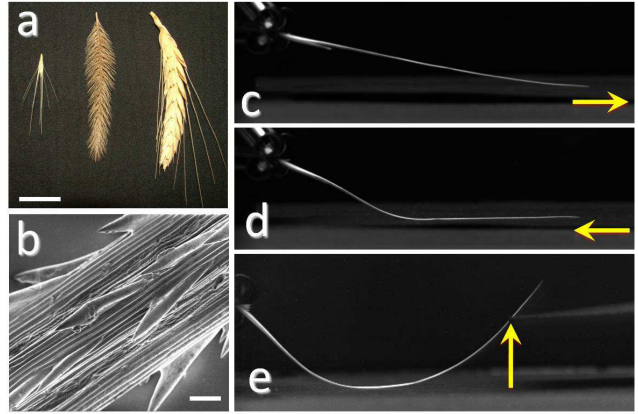}\caption{ (Color online) a) Common
grass species exhibiting ratchet motility (from left to right): single
spikelet of foxtail grass (\textit{Hordeum murinum}), \ \textit{Setaria
viridis} and barley (\textit{Hordeum vulgare}), scale bar 2 $cm$. b)
\textit{Hordeum murinum}'s micro-barb surface structure responsible for
ratchet effect (SEM), scale bar 50 $\mu m$. c) A Single clamped unstrained
\textit{Hordeum murinum} awn (with micro-barbs pointing to the right) glides
easily on substrate moving to the right (paper cover). d) Awn buckling and
progressive micro-barb locking induced by substrate motion to the left. e) A
barb-locked awn on substrate resists also upward lift forces.}%
\label{Fig1}%
\end{figure}

When a \textit{Hordeum murinum} plant awn is placed on a soft, porous or rough
material and excited by macroscopic forces with a variety of origins, it
generically moves in the direction of its tip. To understand the mechanism of
operation of this simple machine, we start with the common observation that
sliding one's fingertips over the surface of a foxtail awn is easy in one
direction and essentially impossible in the opposite direction. A closer look
under a scanning electron microscope (SEM) reveals the reason (Fig 1b). The
awn has a highly anisotropic surface structure with sharp micro-barbs tilted
unidirectionally at an angle $\beta\approx35^{\circ}$ with the horizontal and
a typical length of $50\mu m.$ Such a surface geometry in combination with the
elasticity of the awn induces a cooperative locking transition of micro-barbs
via a progressive bending of the awn toward the substrate (Fig 1d ). This
cooperativity of locking increases the number of stress bearing micro-barbs
and thus reduces the probability of interfacial failure. The tilted barb has a
second effect in addition to providing a force parallel to the substrate,
since its geometry naturally also gives rise to an effective normal
(adhesion-like) force component $F_{\perp}=F_{\parallel}\tan\beta$ as shown in
Fig 1e. The combination of a rigid microscopic geometry associated with the
barbs and a macroscopic elastic compliance associated with the backbone to
which the barbs are attached allows the foxtail awn to slide unhindered with
respect to the substrate in one direction (Fig 1c) but efficiently suppresses
any motion in the opposite direction (Fig 1d). We consider now two different
physical situations in which transport is generically observed.

\begin{figure}[ptb]
\includegraphics*[width=8.5cm]{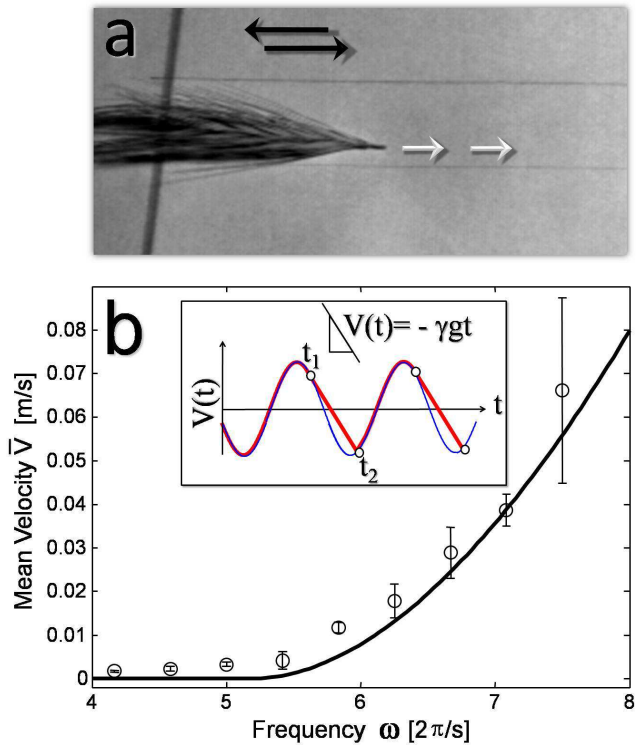}\caption{ (Color online) a) Foxtail
grass moves unidirectionally on an harmonically oscillating horizontal shaker
(cf. supplementary movie 1). b) The mean velocity is shown as a function of
the shaker frequency, with error-bars from 5 different foxtail specimen
(circles). The best fit to a theoretical model (see text) with $\gamma=0.4$ is
shown as a thick line. Inset: Graphical model for ratchet motion on
oscillating substrate. The ratchet leaves the substrate frame when its
inertial force overcomes the friction force at $t=t_{1}$ and it locks and
moves with the substrate again when ratchet and substrate velocities coincide
again at $t=t_{2}$. }%
\label{Fig2}%
\end{figure}

\textit{Transport in inertially dominated environments.} In a
first experiment we placed \textit{Hordeum murinum} spikelets on a
horizontal shaker driven sinusoidally with amplitude $A=3.7mm$ and
observe their motion as a function of the driving frequency
$\omega$. There is a critical frequency $\omega_{c}$ below which
almost no motion is observed. Above the critical shaking frequency
we observe the onset of strong spikelet drift in the direction of
its tip, as shown in Fig 2 (See supplementary movie 1
\cite{SupplMovieShaker}). The foxtail cannot move unless inertial
forces overcome friction, so that we
expect the onset of motion when $\omega>\omega_{c}$ with $\omega_{c}%
^{2}mA=\gamma mg$, i.e. $\omega_{c}=\sqrt{\gamma g/A}.$ Here $m$ is the mass
of the grass, $\gamma$ its friction coefficient and $g$ the acceleration due
to gravity. Above this critical frequency, the mean transport velocity can be
derived from the following graphical argument (cf. inset of Fig 2b): the grass
moves with the shaker up to the point $t_{1}$ where the shaker acceleration
$\dot{v}_{s}\left(  t\right)  $ coincides with the critical slope $-\gamma g$.
From this point on the grass unbinds from the shaker and slides on it,
decelerating under the influence of friction with $\dot{v}\left(  t\right)
=-\gamma g$. The free motion ceases when the grass and substrate velocities
coincide again at a later time $t=t_{2}$ and the cycle begins again. In
general the time-averaged grass velocity can be written as $\bar{v}=\frac
{1}{T}\int_{t_{1}}^{t_{2}}\left(  v_{s}-\gamma gt\right)  dt$. Evaluating this
expression for a sinusoidal shaker velocity $v_{s}\left(  t\right)
=A\omega\sin\omega t$ shows that this agrees well with our experimental
results, with $\gamma=0.4$ (Fig 2b), which value is close to the
experimentally measured friction coefficient of $\gamma=0.46\pm0.05$ (from
independent inclined plane measurements on 6 different \textit{Hordeum
murinum} samples). Close to the transition frequency $\omega_{c},$ a Taylor
expansion for $t_{1}$ and $t_{2}$ in the small parameter $\alpha
=\sqrt{1-\omega_{c}^{2}/\omega^{2}}<<1$ gives simple expressions $t_{1}%
\approx\frac{T}{2}-\sqrt{2}\alpha/\omega_{c}$ and $t_{2}\approx\frac{T}%
{2}+2\sqrt{2}\alpha/\omega_{c},$ leading to $\bar{v}=\frac{1}{T}\int_{t_{1}%
}^{t_{2}}\left(  v_{s}-\gamma gt\right)  dt\approx\frac{9}{2}\Theta\left(
\omega-\omega_{c}\right)  A\omega\left(  1-\left(  \omega_{c}/\omega\right)
^{2}\right)  ^{2}$ with $\Theta$ the Heaviside function. In the other limit of
large frequencies $\omega>>\omega_{c}$ the grass moves almost freely obtaining
momentum from the shaker during a very brief locking phase, so that we expect
an asymptotic mean velocity $\bar{v}\approx A\omega$. Summarizing, the
transport velocity in this frictional-inertial regime is therefore independent
of grass mass and length but obeys a non-linear behavior at onset that
transitions to a simple linear relation far from criticality.

\textit{Transport in continuously deforming environments.} We now turn to the
second, biologically more relevant, situation in which net transport is
observed: continuous deformation of the substrate along which the grass is
moving. Two important examples for such deformations are: a)\ Deformation of
the skins of animals (naturally dispersing the foxtail) b) Diurnal and
seasonal soil swelling and shrinkage driven by humidity and temperature
variations \cite{SoilStrains},\cite{SoilStrains2}. \ The first process is
believed to be responsible for the dispersal of foxtail over large distances.
It poses a hazard to animals if the foxtail is trapped in nostrils or ear
channels, eventually ratcheting through the tissue causing infection and
possibly death. The second process is a potential mechanism for the seed and
awn burial into the soil \cite{OldPaperOnStipa}-\cite{Elbaum}, an important
precursor of germination. \ In general in addition to substrate deformation
the contour length\ of the ratchet grass itself can actively vary as well. For
example, relatives of the foxtails like \textit{Stipa somata} show a
hygroscopically driven coiling and kinking of their awns to vary the effective
contour length of awns (along the direction of motion) which yields linear
strains of up to 20\% \cite{OldPaperOnStipa}. In either case, these strain
variations may be rectified in the direction of the grass tip. To understand
the minimal properties of ratchet propulsion in this case, we consider the
combination of a time-periodic hygroscopic strain of the grass awn
$\varepsilon_{g}\left(  t\right)  $, and that of the surrounding substrate
(soil/animal fur) $\varepsilon_{s}\left(  t\right)  $. Then, on dimensional
grounds, we expect a simple scaling relation of the mean propulsion velocity
of the awn:
\begin{equation}
\bar{V}_{g}\sim aT^{-1}f\left(  \varepsilon_{s},\varepsilon_{g}\right)
\end{equation}
where $a$ is the awn length, $T$ the period of the soil and awn strain
variation and $f\left(  \varepsilon_{s},\varepsilon_{g}\right)  $ a
dimensionless functional.

For a more careful analysis we parameterize the grass using its unstrained
arclength $s_{g}\in\left[  -a/2,a/2\right]  $ with $s_{g}=0$ corresponding to
its center. In response to internal and substrate strain, each point along the
grass moves so that its position is $x_{g}\left(  s_{g},t\right)
=X_{g}\left(  t\right)  +s_{g}\left(  1+\varepsilon_{g}\left(  t\right)
\right)  $ and its velocity is $v_{g}\left(  s_{g},t\right)  =\dot{X}%
_{g}\left(  t\right)  +s_{g}\dot{\varepsilon}_{g}\left(  t\right)  $ with
$X_{g}\left(  t\right)  =x_{g}\left(  0,t\right)  $ and $v_{g}(s,t)=\dot
{x}_{g}(s,t)$. The substrate (soil tube) can be parameterized in a similar
manner via its unstrained arclength coordinate $s_{s}$ so that material points
along it have a position $x_{s}\left(  s_{s},t\right)  =\left(  1+\varepsilon
_{s}\left(  t\right)  \right)  s_{s}$ and velocity $v_{s}\left(
s_{s},t\right)  =s_{s}\dot{\varepsilon}_{s}\left(  t\right)  $. At a given
instant of time $t$ the grass is in contact with the soil over the interval
$x_{s}\in\left[  X_{g}\left(  t\right)  -\frac{a}{2}\left(  1+\varepsilon
_{g}\left(  t\right)  \right)  ,X_{g}\left(  t\right)  +\frac{a}{2}\left(
1+\varepsilon_{g}\left(  t\right)  \right)  \right]  $. Then, the relative
velocity between the grass and the substrate parameterized in terms of the
grass material parameter $s_{g}$ is $V_{rel}\left(  s_{g},t\right)
=v_{g}\left(  s_{g},t\right)  -v_{s}\left(  s_{s}\left(  s_{g},t\right)
,t\right)  .$ We assume that the grass acts as an ideal ratchet , i.e. that no
part of it can move in the \textit{negative} direction (direction of the
micro-barb sharp tip orientation) with respect to the contact points with the
substrate. Then the relative velocity has to satisfy the following physical
condition:
\begin{equation}
\min_{s_{g}\in\left[  -\frac{a}{2},\frac{a}{2}\right]  }V_{rel}\left(
s_{g},t\right)  =0
\end{equation}
Using the fact that $x_{s}\left(  s_{s},t\right)  =x_{g}\left(  s_{g}%
,t\right)  $ and $\min_{s_{g}\in\left[  -a/2,a/2\right]  }\left(
s_{g}C\right)  =-\frac{a}{2}\left\vert C\right\vert $ (for any value $C$) the
upper condition can be recast as an equation of motion%
\begin{equation}
\frac{d}{dt}\frac{X_{g}}{1+\varepsilon_{s}}=\frac{a}{2}\frac{\left\vert
\left(  1+\varepsilon_{s}\right)  \dot{\varepsilon}_{g}-\left(  1+\varepsilon
_{g}\right)  \dot{\varepsilon}_{s}\right\vert }{\left(  1+\varepsilon
_{s}\right)  ^{2}}%
\end{equation}
Integrating the previous expression yields the position of the center of the
awn $X_{g}\left(  t\right)  =\frac{a}{2}\left(  1+\varepsilon_{s}\right)
\int_{0}^{t}\left(  1+\varepsilon_{s}\right)  ^{-2}\left\vert \left(
1+\varepsilon_{s}\right)  \dot{\varepsilon}_{g}-\left(  1+\varepsilon
_{g}\right)  \dot{\varepsilon}_{s}\right\vert dt^{\prime}$. To determine the
mean velocity of the awn, we assume that both $\varepsilon_{s}\left(
t\right)  $ and $\varepsilon_{g}\left(  t\right)  $ are simple periodic
functions with a single minimum and maximum, with same period $T$ and that
$\varepsilon_{s}$ satisfies $\varepsilon_{s}\left(  0\right)  =\varepsilon
_{s}\left(  T\right)  =0.$ Then the velocity averaged over a complete cycle
$\bar{V}_{g}=\left(  X_{g}\left(  T\right)  -X_{g}\left(  0\right)  \right)
/T$ is
\begin{equation}
\bar{V}_{g}=\frac{a}{2T}\int_{0}^{T}\frac{\left\vert \left(  1+\varepsilon
_{s}\right)  \dot{\varepsilon}_{g}-\left(  1+\varepsilon_{g}\right)
\dot{\varepsilon}_{s}\right\vert }{\left(  1+\varepsilon_{s}\right)  ^{2}}dt
\label{VgGeneral}%
\end{equation}

\begin{figure}[ptb]
\includegraphics*[width=8.5cm]{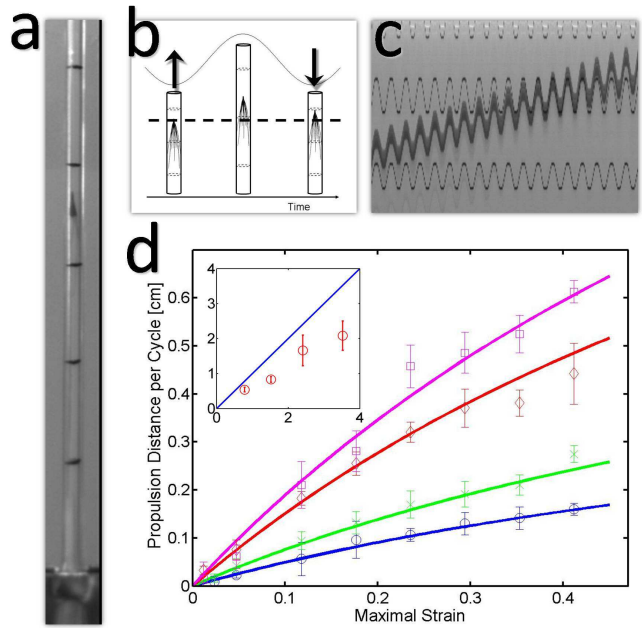}\caption{ (Color online) Ratchet grass
propulsion in a continuously deforming environment. a)-b) Experimental setup:
foxtail spikelets are placed inside a 4 mm inner diameter rubber tube that is
periodically strained (see supplementary movie 2). Spikelets move in the
direction of their tips (stained black for video analysis) with efficiency
varying with their length. c) An intensity line-scan (kymograph) corresponding
to the time dynamics in a) showing the periodic tube motion and the average
propulsion of the spikelet over the time course of 60 seconds (x-axis). d)
Propulsion distance per strain cycle vs. maximal rubber tube strain for 4
spikelets of various lengths: Experimental points with error bars from 7-9
measurements and best fits to $\Delta_{g}^{II}=a\varepsilon_{s}^{\max}/\left(
1+\varepsilon_{s}^{\max}\right)  $ (thick lines). Inset: The fitted effective
contact length (y axis, units of cm) vs. the measured length of the longest
awn (x axis, units of cm) of individual spikelets. While the effective lengths
show a good correlation with the measured lengths, they are slightly smaller
due to occasional awn buckling and imperfect tube-awn contact.}%
\label{Fig3}%
\end{figure}

There are two interesting limiting cases worth considering: \textit{Case I:
Negligible substrate strain.} In this case $\varepsilon_{s}=0$ so that
(\ref{VgGeneral}) simplifies to $\bar{V}_{g}=\frac{a}{2T}\int_{0}^{T}%
\dot{\varepsilon}_{g}dt=\frac{a}{2T}(\int_{0}^{T_{\max}}\dot{\varepsilon}%
_{g}dt-\int_{T_{\max}}^{T}\dot{\varepsilon}_{g}dt)=\frac{a}{T}\varepsilon
_{\max},$ where $T_{\max}$ is the point at which $\varepsilon_{g}$ attains its
maximum and $\dot{\varepsilon}_{g}$ changes sign from positive to negative.
Then the propulsion distance per cycle
\begin{equation}
\Delta_{g}^{I}=T\bar{V}_{g}=a\varepsilon_{\max}
\label{DisplacementGGrassStretches}%
\end{equation}
with $\varepsilon_{\max}$ being the maximal strain during a single strain
cycle. \textit{Case II: Negligible plant awn strain. }In this case we have
$\varepsilon_{g}=0$ and $\bar{V}_{g}=\frac{a}{2T}\int_{0}^{T}\left[
\left\vert \dot{\varepsilon}_{s}\right\vert /\left(  1+\varepsilon_{s}\right)
^{2}\right]  dt$. Splitting the integral as in the previous case, we can
perform the integration over a full cycle and obtain $\bar{V}_{g}=\left(
a/T\right)  \varepsilon_{\max}\left(  1+\varepsilon_{\max}\right)  ^{-1}$ and%
\begin{equation}
\Delta_{g}^{II}=a\frac{\varepsilon_{\max}}{1+\varepsilon_{\max}}
\label{DisplacementGSubstrateStretches}%
\end{equation}
We see that the displacement per cycle $\Delta_{g}$ scales in both cases
linearly with the grass length $a$ showing that longer grasses pick up and
rectify larger fluctuations in proportion with their length. Therefore longer
grass awns act effectively as better "fluctuation amplifiers"\textbf{.}

To test our results quantitatively, we focus on the case when substrate
strains dominate, since this is simple to setup experimentally. To simulate
the behavior of a ratcheting awn in contact with soil, awns and single rachis
spikelets of various grasses of different lengths and diameters (Fig1 a) were
placed inside a rubber tube (inner diameter 4 mm) . The tube was then
subjected to slow periodic strains with maximal strains ranging from
$\varepsilon^{\max}=0.01$ to $0.4$. All the grasses rectified the oscillatory
strain with the \textit{Hordeum murinum} spikelets showing most robust and
reproducible behavior (Fig 3 d). At very small sub-Hertz frequencies at which
the rubber tubes were strained (0.3 Hz) the ratchet displacements per cycle
were independent of both frequency and the direction of gravity. The
experimental results for the propulsion displacement $\Delta_{g}$ as a
function of the length of \textit{Hordeum murinum} spikelets show good
agreement with the theoretical prediction
\ref{DisplacementGSubstrateStretches} with a single fitting parameter per
specimen - the effective contact length $a$, which itself is proportional to
the actual awn length (Fig 3 d, inset). For longer awns the fitted length
shows however a systematic 20-40\% reduction relative to the measured length
consistent with our observation of occasional awn buckling and imperfect tube
- awn contact of longer spikelets; this effect is less pronounced for shorter
spikelets (Fig 3 d, inset).

Interestingly the results (\ref{DisplacementGGrassStretches}%
)-(\ref{DisplacementGSubstrateStretches}) are consistent with the previously
measured linear relation between the awn length and burial depth of
\textit{Stipa somata} \cite{StipaLengthVsSpeed} and provide some evidence for
a selective advantage of larger awn lengths. More generally, for
hygroscopically driven awns (\ref{VgGeneral}) predicts a non-trivial
interaction between soil strain and awn strain contributions to the
rectification of motion. Indeed the two effects act in concert or "resonate"
for oppositely phased strains but can nullify each other for in-phase strains.

Our simple experiments and theory allow us to understand the physics of plant
awn ratchets. Equipped with anisotropically toothed microstructured elastic
awns the seeds are able to lock directionally into virtually any soft natural
material, like soil, animal fur and skin. The pronounced asymmetry of their
surface structures enables the seeds to rectify external environment
fluctuations along the direction imposed by the specific orientation of their
micro-barbs. Furthermore, longer awns profer a simple selective advantage to
those seeds since they can move further and thus disperse more efficiently
than their shorter competitors. Our results also point the way for
technological designs that might mimic this simple but robust motif.


\begin{thebibliography}{99}                                                                                               %


\bibitem {Feynman}Feynman Lectures on Physics, Addison-Wesley, Vol. 1, 1964.

\bibitem {Ajdari}Ajdari A and J.R. Prost, C. R. Acad. Sci. II (Paris), 315,
1635 (1992)

\bibitem {Magnasco}M.O. Magnasco, Phys. Rev. Lett. 71, 1477 (1993)

\bibitem {BrownianRatchet}P. Reimann, Phys. Rep. 361, 57 (2002)

\bibitem {Ratchet}F. Julicher, A. Ajdari A, J.R. Prost, Rev. Mod. Phys. 69,
1269 (1997)

\bibitem {Kulic}I.M. Kuli\'{c}, R.M. Thaokar and H. Schiessel, Europhys. Lett.
72, 527 (2005)

\bibitem {BugBerne}A. Bug and B. Berne, Phys. Rev. Lett. 59, 948 (1987)

\bibitem {MacroscopicRatchetDevice}B. Nord\'{e}n, Y. Zolotaryuk, P. L.
Christiansen, and A. V. Zolotaryuk , Phys. Rev. E 65, 011110 (2001)

\bibitem {Mahadevan}L. Mahadevan, S. Daniels and M. Chaudhury, Proc. Natl.
Acad. Sci. (USA), 101, 23 (2003).

\bibitem {OldPaperOnStipa}L. Murbach, Botanical Gazette, 30, 113 (1900)

\bibitem {SeedBurialPeart}M.H. Peart, J. Ecology, 67, 843 (1979)

\bibitem {SeedBurialStamp}N.A. Stamp, J. Ecology, 72, 611 (1984)

\bibitem {StipaLengthVsSpeed}L.K. M. Garnier and I. Dajoz, Ecology, 82,1720 (2001)

\bibitem {Elbaum}R. Elbaum, L. Zaltzman, I. Burgert and P. Fratzl, Science
316, 884 (2007)

\bibitem {SoilStrains}S. Iwata, T. Tabuchi, B. P. Warkentin, Soil-Water
Interactions, M. Dekker New York and Basel (1988).

\bibitem {SoilStrains2}Soil Physics, T. J. Marshall, J. W. Holmes, and C. W.
Rose, Cambridge University Press (1996).

\bibitem {SupplMovieShaker}Online supplementary movie 1: Hordeum murinum
moving on a laboratory shaker above the critical frequency. Recorded at 200
fps (frames/s) , reproduced at 25fps.

\bibitem {SupplMovieTube}Online supplementary movie 2: Hordeum murinum
spikelet moving in a stretching rubber tube. Recorded at 5 fps and reproduced
at 25 fps. Color mark distance on tube in relaxed state = 2.54 cm.
\end{thebibliography}
\end{document}